\documentclass[sigconf]{acmart}

\AtBeginDocument{%
  \providecommand\BibTeX{{
    \normalfont B\kern-0.5em{\scshape i\kern-0.25em b}\kern-0.8em\TeX}}}
\usepackage{multirow}
\setcopyright{rightsretained}

\usepackage{fancyhdr}
\copyrightyear{2020}
\acmYear{2020}
\acmConference[CHIIR '22]{ACM SIGIR Conference on Human Information Interaction and Retrieval}{July 03--05, 2018}{Woodstock, NY}
\acmBooktitle{ACM SIGIR Conference on Human Information Interaction and Retrieval,
  June 03--05, 2018, Woodstock, NY}
\acmPrice{15.00}
\acmISBN{978-1-4503-XXXX-X/18/06}
\begin{document}


\title{
Watch Less and Uncover More:\\
Could Navigation Tools Help Users Search and Explore Videos?
}





\author{Mar\'ia P\'erez-Ortiz}
\email{maria.perez@ucl.ac.uk}
\affiliation{
    \institution{University College London}
    \city{London}
    \country{United Kingdom}
}
 \authornote{First and second author contributed equally to this research.}
\author{Sahan Bulathwela}
\email{m.bulathwela@ucl.ac.uk}
\affiliation{
    \institution{University College London}
    \city{London}
    \country{United Kingdom}
}


\author{Claire Dormann}
\email{c.dormann@ucl.ac.uk}
\affiliation{%
  \institution{University College London}
  \country{United Kingdom}}

\author{Meghana Verma}
\email{vermameghana1299@gmail.com}
\affiliation{%
  \institution{Indian Institute of Technology Bombay}
  \country{India}
}

\author{Stefan Kreitmayer}
\email{s.kreitmayer@ucl.ac.uk}
\affiliation{%
 \institution{University College London}
 \country{United Kingdom}}

\author{Richard Noss}
\email{r.noss@ucl.ac.uk}
\affiliation{%
  \institution{University College London}
  \country{United Kingdom}}

\author{John Shawe-Taylor}
\email{j.shawe-taylor@ucl.ac.uk}
\affiliation{%
  \institution{University College London}
  \country{United Kingdom}}

\author{Yvonne Rogers}
\email{y.rogers@ucl.ac.uk}
\affiliation{  \institution{University College London}
  \country{United Kingdom}}

\author{Emine Yilmaz}
\email{emine.yilmaz@ucl.ac.uk}
\affiliation{  \institution{University College London}
  \country{United Kingdom}}

\renewcommand{\shortauthors}{P\'erez-Ortiz et al.}

\begin{abstract}
Prior research has shown how ‘content preview tools’  improve  speed  and  accuracy  of  user relevance  judgements  across different information  retrieval  tasks. 
This paper describes a novel user interface tool, the Content Flow Bar,  designed to allow users to quickly identify relevant fragments within informational videos to facilitate browsing, through a cognitively augmented form of navigation. It achieves this by providing semantic “snippets” that enable the user to rapidly scan through video content. The tool provides visually-appealing pop-ups that appear in a time series bar at the bottom of each video, allowing to see in advance and at a glance how topics evolve in the content. 
\textcolor{black}{
We conducted a user study to evaluate how the tool changes the users search experience in video retrieval, as well as how it supports exploration and
information seeking. The user questionnaire revealed that participants found the Content Flow Bar helpful and enjoyable  for finding relevant information in videos.  The interaction logs of the user study, where participants interacted with the tool for completing two informational tasks, showed that it holds promise for enhancing discoverability of content both across and within videos. This discovered potential could leverage a new generation of navigation tools in search and information retrieval.}
\end{abstract}

 \ccsdesc[500]{Information systems~Users and interactive retrieval}
 \ccsdesc[300]{Information systems~Search interfaces}
 \ccsdesc[100]{Information systems~Information extraction}
 \ccsdesc[500]{Applied computing~Interactive learning environments}

\keywords{Video retrieval, user interfaces, interaction design, navigation, search, open education}

\maketitle

\section{Introduction}

One of the primary bottlenecks in information retrieval tasks is that users can only access content of a retrieved item serially and sequentially. This not only applies to processing multiple materials, but also, importantly, to filtering lengthy ones, since users may not know at the outset which  parts  of the content (e.g. a video) might be relevant to their needs. To this end, ‘content previews’
could improve  the  speed  and/or  accuracy  of  user relevance  judgements  across  a  variety  of  content  types  and  information  retrieval  tasks \cite{query_biased_IR}. 
However, much of the potential to build better navigation tools still appears to be unused in search and retrieval tools. \textcolor{black}{Analyses of how these tools affect user behaviour, perception and search experience are also scarce.} 


Written documents, such as ebooks, academic papers and lecture notes, constitute a substantial fraction of educational resources. While these are generally considered to be core learning resources, research in online learner behaviour has shown that it can be overwhelming and unwieldy to use them in practice, often being skimmed over \cite{Guo_vid_prod}. An increasingly popular approach is for students to watch video lectures that are provided freely online. 
Students find them engaging and more accessible \cite{lan2017behavior}. However, the number of educational online videos that are available has grown exponentially, making it difficult for users to choose. While they can use a search engine in YouTube or Google, this typically provides thousands of potential videos based on metadata such as title and description. Do they select the top one on the list, which may be peripherally on topic, not at the right level, or even boring? Is there another way to provide a way of helping users choose? 

In this work, we propose and evaluate a navigation tool to support learners and teachers choose from the potentially millions of videos available and access relevant educational content.
Our approach is to design a visualisation tool that highlights fragments of videos that match the search query and can serve as effective entry points (or alternatives - depending on the learner's information needs at hand) rather than starting from the beginning of a video. The goal is to increase transparency, reduce user effort and put the learner in control of their choices \cite{10.1007/s11042-016-3661-2}.


 \textcolor{black}{Our focus is on evaluating the tool (named Content Flow Bar, CFB) for educational videos in terms of how it affects user behaviour and informational search experience.}
To evaluate the tool, we conducted a user study with two information seeking tasks. Participants were asked to use the tool to decide which videos best matched each task question. Interaction logs and qualitative survey results were collected \cite{deepest_rank}. Our results show that Content Flow Bar is highly useful in helping users navigate content, with $82\%$ of the participants agreeing that CFB made finding video clips easier and approximately $93\%$ of the participants agreeing that more video players should include the CFB. The interaction logs further showed that the CFB tool leads to less time watching content, more exploration both within and across videos and less dwell time per opened video. Participants also navigate deeper and jump (seek) less within the content. Finally, participants take less time before making a relevance judgement. We hypothesize this may be the result of users being able to follow different paths to find what best matches their search, as well as eliminate irrelevant information.


\section{Related Work}

\textcolor{black}{Our proposed tool attempts to improve efficient previewing and non-linear consumption of videos with intelligent user interface components and artificial intelligence. In this section, we identify several relevant contributions in information retrieval (IR, section 2.1) and video retrieval (section 2.2) that have inspired our work.}

\subsection{Information Search and Retrieval}

\textcolor{black}{The key objective of designing IR systems is to facilitate the task of users of finding relevant information searching inter-/intra- documents. With the distinction between search and recommendation blurring by the day \cite{blurring}, making results personalised, interactive and insightful is becoming crucial. Users primarily rely on textual and visual cues such as title, snippets (query dependent textual summaries \cite{10.1155/2018/7836969}), descriptions and thumbnails during the search interaction process. Commercial search engines also use techniques such as highlighting query words \cite{10.1155/2018/7836969}. The proposed tool leverages multiple textual queues to incorporate these learnings.}

\textcolor{black}{Increased transparency, humanly-intuitive facets, and detailed  metadata/summaries have all been proven to consistently help  the  user  in  selecting material \cite{anderson2019strategies}. For example, efficient summaries of search results have enabled users to access deeper ranked documents, i.e., enhance exploration \cite{10.1145/2484028.2484084, deepest_rank}. 
Cortinovis et al. \cite{cortinovis2019supporting} is another example showing success, where the researchers developed Google-like IR search interfaces incorporating supplementary information such as document's alignment to the keyword and similarity-based grouping of documents. 
This work shows that in educational IR settings, such a blended search experience may be of special importance, as users may not be entirely familiar with the topics they search for (since they are trying to learn about them). This means that in an educational scenario, allowing users to expand on search results, in order to understand the content better, may be preferable. 
} 

\textcolor{black}{Having efficient previewing capabilities has also been proven instrumental to user relevance judgments in search. From generating static abstracts of texts (descriptions) \cite{auto_abstract_index}, to more recent query-based summaries, both cases have been shown to improve speed and accuracy of result relevance by “helping users to more easily identify the relevant pieces of information that are contained in each document” \cite{query_biased_IR}. When previewing information, methods such as \emph{WordClouds} (a cluster of words) \cite{williams2014finding} and \emph{TileBars} (a heatmap like representation of word presence) \cite{tilebars} enable fast visual representations (that use size, color, etc.) by reducing textual information and emphasising ranks of keywords (size of word in WordCloud and colour intensity in TileBar). While our approach uses concept annotation for summarising content, it also ranks concepts to indicate their authority within a video fragment, with the objective of leveraging speed and accuracy of relevance judgments.}

\begin{figure*}[ht!]
\centering
\includegraphics[width=\textwidth]{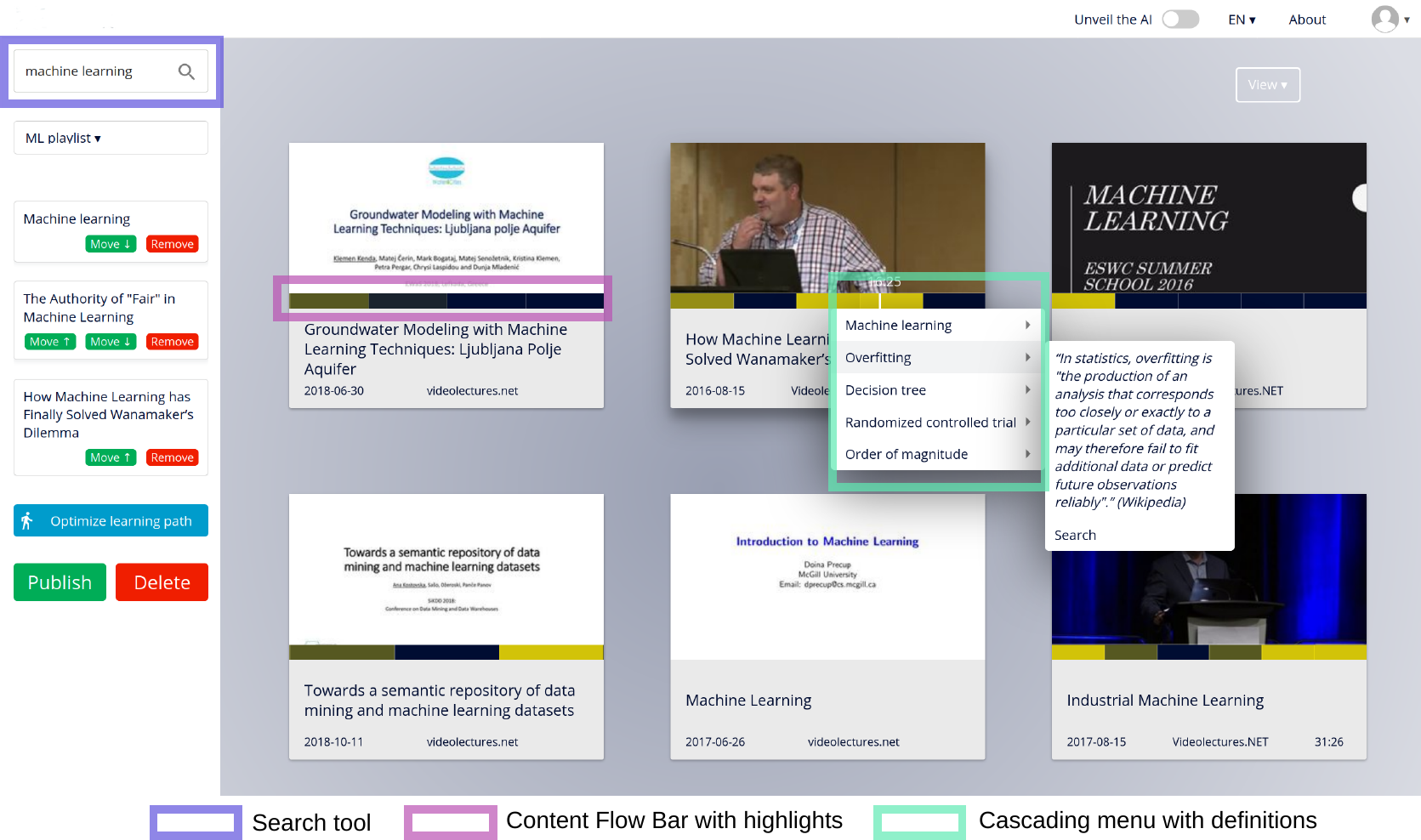}
\caption{Platform in which the CFB was integrated (X5Learn), allowing to access both videos and PDF documents as search results. The blue box shows the search bar. The purple box shows the CFB with relevant content highlighted in shades of yellow. The last box shows the CFB cascading menu, which shows the topics covered in the video and allows to find the definition of those. \textcolor{black}{Note that many features of the platform were disabled for the user study, for example participants did not have access to relevance highlights (marked in yellow in the CFB), since searching as a feature was disabled. } }
\label{fig:platform}
\end{figure*}

\subsection{Judging Relevance in Informational Videos}

\textcolor{black}{We analyse now the specific case of informational video retrieval. Navigation within videos is not as straightforward as text, since information is embedded within video frames meaning it would often be consumed sequentially. However, side information, such as title and description, is commonly used to create semantically related thumbnails when previewing contents in videos \cite{deep_thumb,vasudevan2017query}. Variants such as grids of animated thumbnails \cite{Jackson2013PanopticonAP} and numerous progress bar-based “skimming” approaches produce previews similar to YouTube or Netflix \cite{10.1145/274644.274670,Justin2013}. Skimming is found to be a common strategy among young adult self-directed learners when previewing educational videos \cite{10.1007/978-3-319-70232-2_27}. While previewing videos is mainly associated with thumbnails alone, research shows that utilising both textual metadata and thumbnails leads to more efficient relevance judgements \cite{dziadosz2002thumbnail}. Our work builds on these findings to use metadata, transcript-based concept annotations and thumbnails to improve efficient previewing and navigation.}

\textcolor{black}{Several techniques have explored utilising annotations of videos to improve IR. Some of these, although promising, use crowd-sourcing to harvest video information such as timed-metadata \cite{DBLP:journals/jifs/PintoV19} and video fragment concept annotations \cite{10.1007/978-3-030-05710-7_12,hierarchicalbrushing}, which is not scalable without a significant workforce. To this end, neural \cite{miech2019howto100m} and semantic-based \cite{yu2016strategies} automatic video representation tools have been proposed for video retrieval tasks. However, none of these techniques have been studied in terms of their usability in video previewing to make relevance judgements. YouTube Chapters (textual titles for segments of video) has also recently automated video fragment annotation. While this tool allows non-linear navigation within a video, our approach allows smart previewing when navigating through video search results, i.e. it can be used before selecting a video. Additionally, YouTube Chapters uses optical character recognition to identify text in video frames, restricting the approach to presentation-like (slide shows) videos\footnote{\url{https://support.google.com/youtube/thread/18138167/youtube-test-features-and-experiments}}. The tool's design and experimental results are not publicly available, restricting the evaluation of its success. }

\textcolor{black}{Contrary to these tools, our idea utilises automatic concept annotation using Wikipedia topics \cite{wikifier}, allowing users to expand their knowledge on unfamiliar topics during search by means of Wikipedia definitions \cite{cortinovis2019supporting} and can work with any type of informational video as it is transcript-based. The user study also provides a much needed usability evaluation of the intelligent video previewing tool proposed, showing the potential of navigation tools in informational video retrieval.}

\section{The visual NAVIGATION Tool}

  
  We present a new visual preview tool to facilitate informational browsing. This interaction technique, which we call the Content Flow Bar (CFB), was designed to provide semantic “snippets” of video content, intended to be useful for both searching and exploring. These snippets are displayed as pop-ups and overlap in a time series bar, allowing the user to see in advance what topics are covered in a video. The rationale for this design  was for it to be a lightweight tool that could provide just enough details that could be glanced at while skimming a video thumbnail to determine if the pop-up keywords match the perceived need. 
We introduce now the CFB, together with the design principles and context of use, as well as the backbone for extracting the pop-up keywords, wikification.

\begin{figure*}[t!]
\centering
\includegraphics[width=\textwidth]{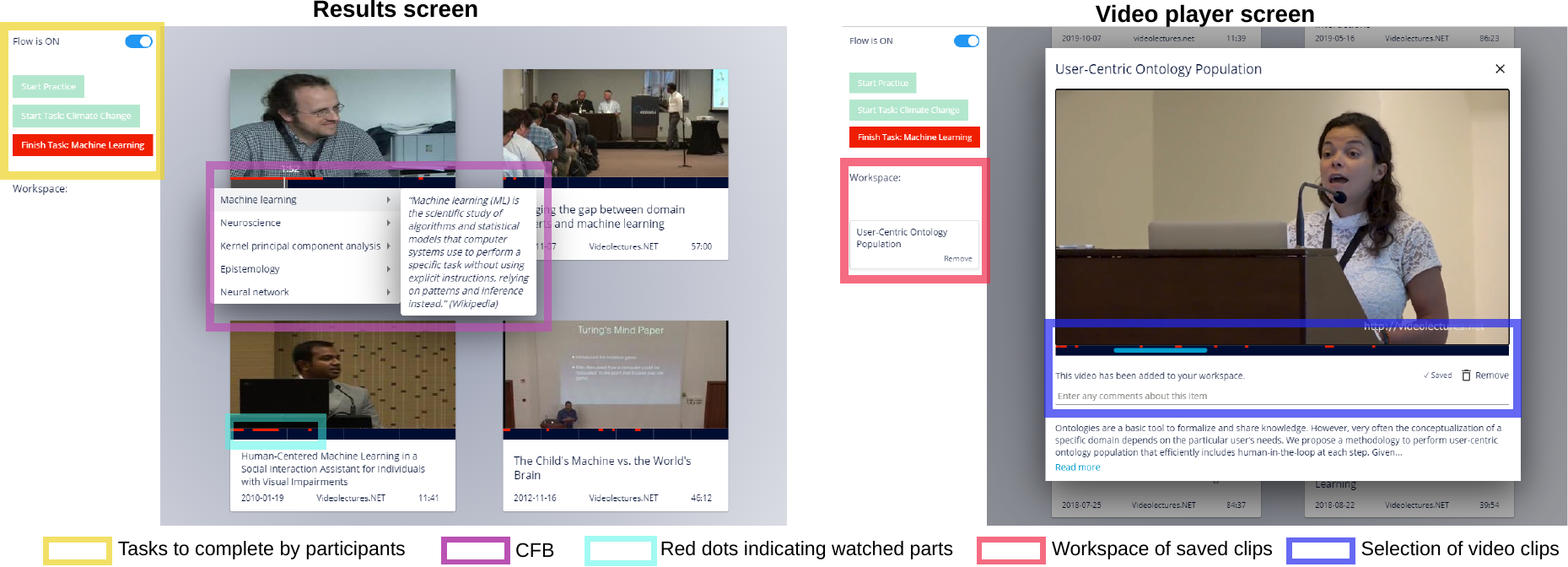}
\caption{Screenshots of the study, highlighting important components. Left sub-figure shows the \emph{Results} screen where the videos are displayed for each task. The right one shows the \emph{Video player} screen (pop-up) that opens when clicking on a video.}
\label{fig:userstudy}
\end{figure*}

\subsection{Context of Use}
The CFB has been deployed in our educational platform\footnote{Demo available with regristration at \url{https://x5learn.org/}}, designed to enable anyone to view online open educational material. \textcolor{black}{One intended function is to enable learners and educators to be able to  construct their own curriculum, e.g. retrieving open educational materials and judging the potential relevance of the content. It was designed for both learners and educators, but here we evaluate its usefulness and usability for the latter group.}
The platform, together with the search tool, CFB and cascading menus, can be seen in \figurename{ \ref{fig:platform}}. This screenshot shows what a user would see when joining the platform and searching for the topic of ‘machine learning’. The resulting videos that the system returns are shown with a thumbnail view. A CFB is integrated with each video, \textcolor{black}{allowing to visualise the (Wikipedia) topics covered in time in a video and  highlighting the most relevant parts of each video with regards to the query. Note that the CFB also includes definitions of Wikipedia concepts, which users can access in case they are not familiar with the term.} To make it intuitive to use, our design leverages familiar UX patterns and techniques, primarily the use of popups, colour-coded widgets and cascading menus. \textcolor{black}{The platform also includes other integrated features/tools that are intended to be used with the CFB. These are described in \cite{x5learn,fragmentbar} and aim to  support the learning experience (e.g. optimization of learning paths, annotation of videos, a recommender system \cite{truelearn,trueeducation}, etc.).    } 


\subsection{Content Flow Bar (CFB)}
 The CFB provides semantic “snippets” of video content that pop up for different video fragments. An initial pilot study revealed  that enabling the user to remain in context was important. Therefore, popups were used to display the video player
 rather than opening a new page when clicking on a video (see right sub-figure in \figurename{ \ref{fig:userstudy})}. 
The cascading menus in the CFB (as highlighted with a green square in Figure~\ref{fig:platform}) show the keywords of the video at that particular timestamp, allowing the user to navigate through the topics covered in the video without having to click on the content. As part of the CFB, keywords’ definitions are also provided in the form of pop-ups. These keywords and their definitions have been automatically extracted from the wikification process described later.  

CFB is further supported by using different colour intensities (yellow, see Figure~\ref{fig:platform}) that indicate the predicted relevance of each fragment to the query.  Combining these familiar visual interface elements was intended to enable the user to rapidly glance through a matrix of thumbnails and choose the most relevant. 
Overall, the type of cueing used in the CFB is intended to enable the user to see at a glance what a video lecture covers and to be able to stop at particular points to discover more. 
From the main window, when a participant wants to look further at a fragment of a video, they can jump directly to that fragment. Then, the video player window opens and the video starts at this particular fragment. All participants can also explore the videos by moving their cursor along the video timeline. 

\subsection{Wikification to Extract Keywords}

Our tool uses transcription and translation algorithms from \emph{TransLectures} project\footnote{\url{www.translectures.eu}} to get a text representation of videos. 
In the Wikification process, the English text representation (translated when needed) of the resources is used to partition long documents into fragments of approximately 5000 characters (approx. 5 minutes in video) as prior work finds such fragmentation to be appropriate \cite{truelearn}. Each fragment is then automatically annotated using the open Wikifier service\footnote{\url{www.wikifier.org}}, {which uses natural language processing entity linking methods to label text with relevant and salient Wikipedia concepts} \cite{wikifier}. In doing so, our approach is domain-agnostic, leveraging Wikipedia which is the largest and ever evolving knowledge base, avoiding the need for expensive expert labelling, while working with multiple languages and  allowing for humanly interpretable annotations that enable explainability and scrutiny \cite{filip_explainable}.
{For more information on wikification, including its performance on entity linking tasks, see \cite{wikifier}.} 

\section{User Evaluation}

During the design of the tool we conducted a preliminary user study (preceding the actual study) with 8 participants who were given two versions of an information retrieval task, with and without CFB. Qualitative feedback indicated that all participants found the tool engaging, intuitive and helpful for finding information in videos. The benefits were especially pronounced with long videos of unknown content. As one participant stated: \emph{"It's very hard if you don't know the video, to know where the content is, where the beginning and the end of one thought are. [...] Without [CFB] I had to watch for a long time before it got to the point I was interested in"}. With CFB, the participant found the task easier and more enjoyable \emph{"because you can skip to the part you are actually interested in"}. 

\subsection{Controlled User Study}
A user study involving 26 participants was conducted to evaluate the efficacy of the CFB for {information retrieval and gathering. }

\subsection{Goal of Study and Research Questions} \label{sec:goals_and_RQs}

\begin{figure}[ht!]
    \centering
    \includegraphics[width=0.33\textwidth]{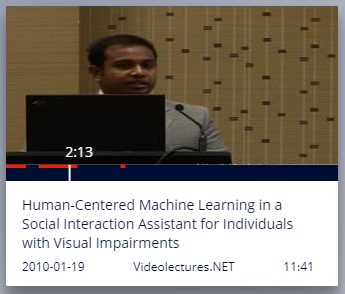}
    \caption{Video player with CFB off. Cursor in the time line indicates the time position. Red dashes and dots indicate watched parts.}
    \label{fig:control}
    \end{figure}

The main user study was designed to compare user performance of a video player with and without the CFB {(control condition). More specifically, the control video player was based on YouTube, showing changing thumbnails when hovering, and highlighting watched parts (see \figurename{ \ref{fig:control}}).  The enhanced video player had the CFB, including a list of topics covered in each fragment and the definition of those (as per \figurename{ \ref{fig:userstudy}}).} This comparative approach is a common method used when evaluating video player interactions \cite{dynamic_slide,anchor_slides}. 

We designed two tasks relating to machine learning and climate change. 
{Machine learning was chosen because of being an academic popular topic. Climate change was chosen because of being of universal concern. 
Both of these are also topical and there are many videos available in the platform where the CFB was integrated.} Both were information seeking tasks in which participants were instructed to find relevant video clips that would be useful for teaching each domain. {We selected an informational task over a navigational one, as informational tasks tend to be more difficult and time consuming. The tasks were open ended and participants could take as much time as needed to complete each task.  This was also because we were interested in measuring the time to complete the task with/without CFB
in an open-ended fashion, so that participants could explore the videos and look up tool in their own time rather than be primed with instructions to complete tasks in a certain time.}

The goal of the study was to understand whether and how the CFB supports information seeking, was able to facilitate content navigation and support efficient browsing of video content. The  two main research questions for the study were as follows: 
\begin{itemize}
    \item How does navigation and browsing compare across the two interfaces?
\item What are the differences in terms of information seeking and exploratory behaviour? 
\end{itemize}


\subsubsection{User Actions} \label{sec:user_actions}
To answer the research questions, {we pay special attention to specific aspects of performance and user experience, focusing on five actions}: (i) Opening videos, i.e. when the participant decides to click on a video and the video player screen is opened; (ii) playing/watching videos in the video player screen; (iii) exploring a video (either in the results screen or the video player screen), which refers to exploring the contents of a video such as keywords in the CFB or frames with the thumbnails; (iv) selection of relevant video segments, i.e. when the user decides that a part of a video is relevant to the task at hand and saves the relevant segment in  the workspace and v) information seeking within a video \cite{seek_link}, i.e. a seek operation is when a user navigates within the video (clicks/jumps on multiple time locations in the video to play).

The actions of opening, playing and selecting parts of videos are the same for both conditions (CFB on and off). However, exploration differs.  
In both cases, exploring the content is done by interacting with the time series bar below the video (present in both the results and video player screens). When CFB is off, the exploration experience is kept similar to that of the mainstream video player in YouTube. This is, hovering over the time series bar in the result screen, will display the time position of the video corresponding to mouse cursor. Additionally, the video thumbnails previewing the frames corresponding to the time position are displayed. In the video screen, the time positions are retained while the thumbnail displaying is disabled. The watched parts of the video are marked in red. \figurename{ \ref{fig:control}} shows a representation of the video player when CFB is off. If the CFB is on, the only difference is that the time is replaced by CFB popups with Wikipedia keywords and definitions. Similar to the CFB off condition, the thumbnail previews are enabled in the results screen and disabled in the video screen. 

\subsection{Experimental Design and Procedure} \label{sec:experiment}

The study used a repeated measures design, so each participant does both tasks with CFB and the control condition, with counterbalancing across them to address training effects and fatigue. \textcolor{black}{This means that the order in which the participants were presented with the two conditions and tasks was randomised.} \textcolor{black}{The CFB condition was demonstrated in the practice task.}


Before the study participants were sent an information sheet and consent form to sign before the session. The study took place through Zoom web conferencing platform. The researcher first gave participants an initial form asking demographics questions and to rate their knowledge of both climate change and machine learning.
knowledge for the topics of the two tasks. Then, the researcher gave a brief overview of the study and demoed the platform, highlighting the features needed for the study (including CFB on and off). The participants then practised with both conditions for the topic “brain” until they felt confident that they could move onto the main study. Then, each participant performed the information seeking task for both topics {(task order randomised). As explained before, no time constraints were given to explore the use of the CFB in an open-ended environment.} After this they were asked to fill in a questionnaire about the CFB.  \textcolor{black}{Being an information gathering task, if users get familiar with the knowledge area over multiple tasks, this will have an impact on the interaction behaviour that is captured and analysed. Three different topics (brain, climate change and machine learning) are used in the three tasks (practice, task 1 and task 2) to the user to avoid this learning effect that might confound the analysis.}




\figurename{ \ref{fig:userstudy}} shows a summary of the user study. For each task (machine learning and climate change), participants were presented in the results screen with 18 videos that were carefully chosen for the study. \textcolor{black}{The video result set shown in \figurename{ \ref{fig:userstudy} (left)} was ranked randomly, i.e., videos were ordered independently of their relevance to the task topic.  Note, however, that all the videos were selected to be relevant for solving the task. This ranking was kept constant among all participants to make sure each participant sees the videos in the same order.} The baseline video player was based on YouTube, showing changing thumbnails when hovering, and highlighting watched parts with red dots. The enhanced videoplayer had the CFB, including a list of topics covered in each fragment and the definition of those.
\textcolor{black}{As the user study is restricted to a static result set (contrary to a user query driven dynamic result set), the relevance-based fragment highlighting feature of CFB was disabled in the user study interface found in \figurename{ \ref{fig:userstudy}}.}
If a participant clicked on a video, then the video player opened taking the participant to the video player screen (see \figurename{ \ref{fig:userstudy}}). Participants were also provided with functionality to select video clips and save them to a workspace. Once a participant has selected a video clip, the selected segment is highlighted in blue on the CFB, and the video title is displayed in the participant’s workspace. Participants could also remove any selection from the workspace. 



\paragraph{Tasks} The participants were presented with the following hypothetical scenarios from which to complete the two tasks:
\textbf{Task 1 (climate change):} “As you are interested in climate change, a friend has asked you to find 2 interesting video clips that illustrate the applications of data science in climate change. After that, your friend wants you to find 2 video clips that illustrate controversial issues in climate change. These video clips will serve to initiate a debate in class. As watching videos should not take too long, you will have to select \textcolor{black}{at least} 4 short video clips or segments for students to watch (e.g. 5-6 minutes). 
\textcolor{black}{Please be advised not to select more than one video clip/segment per video.”}

\textbf{Task 2 (machine learning):}  “Some students want to learn about machine learning in your next workshop. As its importance is growing, you are tasked with finding video clips that illustrate key concepts that students can watch at home, and make notes about. Then, in the workshop, they will like you to show 2 video clips on the applications of machine learning, and discuss these with them. 
As watching videos should not take too long, you will have to select \textcolor{black}{at least} 4 short video clips or segments for students to watch (e.g. 5-6 minutes). 
\textcolor{black}{Please be advised not to select more than one video clip/segment per video.”}


\paragraph{Videos} All the videos selected for the study are found on the platform and belong to the repository VideoLectures.net. We selected 18 educational videos from each subject. All the videos are from academic conference presentations (or specialised seminars and summer schools). In most cases, the speakers’ PowerPoint presentation has been captured and included in the videos. We made sure that the tasks were achievable given the selected set of videos.

\paragraph{Questionnaire} 
 We developed a questionnaire about specific aspects of the CFB to understand user experiences with the tool. The questionnaire contained both open-ended questions and ratings \textcolor{black}{that are inspired by the System Usability Scale \cite{jordan1996usability}}. The questionnaire started with questions about demography and teaching experience. We also requested participants to self-report their knowledge of each topic. 
\textcolor{black}{This part of the questionnaire relating to demography and knowledge was done prior to participants interacting with the tool. The second part of the questionnaire, which included questions relating to the CFB, was asked after participants took part in the study.} \textcolor{black}{In the second part of the questionnaire, }
participants were asked first  to rate statements about the CFB, and then more specifically to rate the adequateness of the topics generated by entity linking and used in the CFB. 
 Some of the statements participants were asked to rate included: i) The enhanced videoplayer (with CFB) is intuitive to use. ii) The enhanced videoplayer made finding videoclips easier. iii) I found information seeking with the CFB difficult. iv) Providing definitions of the topics was helpful. v) More videoplayers should include the CFB.  
 We used 6-item Likert-scales, from strongly disagree to strongly agree.
Finally, participants were asked to describe in detail their experience with the CFB, the user study and the keywords generated.
To conclude, participants were asked if they had any additional comments related to the study, tool or tasks. \textcolor{black}{A copy of the used questionnaire is attached with the supplementary materials.}

 \paragraph{Qualitative analysis}
 \textcolor{black}{We conducted an inductive thematic analysis, using the answers to open-ended questions in the study and the interview transcriptions. The goal of the thematic analysis is to better understand the experiences, perceptions and reactions of participants to the CFB. We coded the data following the principles of thematic analysis outlined by Braun and Clarke \cite{braun2006using}. We followed the main steps inherent in this method, including coding and review of the codes, grouping them into categories and then defining themes. At each stage we strove to  ensure that the emerging patterns in the analysis came from the available data and reflected the experiences of our participants.}




\subsection{Participants}

26
participants took part in the study. Most of the participants were aged between 30-39 years. {They were all staff (PhD students, postdocs and faculty) from the Department of Computer Science at i) University College London and ii) The Open University (UK).  They all had teaching experience and were computer literate.} We had slightly more male participants (53.6\%) than female (46.4\%). Their knowledge of the two topics - machine learning and climate change - varied. Most participants reported having average or above knowledge of machine learning, 40\% had specialist knowledge in machine learning. By contrast, 50\% of participants had little knowledge, while the rest had mostly an average knowledge, of climate change. \textcolor{black}{Note that while we did not explicitly test for prior knowledge} \textcolor{black}{(but, rather allowed self-reporting)}, \textcolor{black}{all of the participants came from computer science departments, thus it is expected that they will have at least basic knowledge of machine learning.}

\subsection{Click Stream and Interaction Log Analysis}

We implemented additional programming snippets within the user interface to capture high detailed user interaction logs related to clicks and other interactions with relevant user interface components (e.g. hovering over the objects, triggering and closing pop-ups, etc.) \cite{mouse_movement}. The recorded usage logs are post-processed to infer interaction patterns by computing different interaction metrics\cite{Guo_vid_prod}. 

\subsubsection{Interaction Metrics}

We calculate different interaction metrics used in measuring user behaviours in IR systems that can be grouped into multiple categories based on time \cite{search_utility,search_task_duration}, activity \cite{search_task_duration,mouse_movement}, location \cite{search_utility,deepest_rank} and selection of video segments. \textcolor{black}{As the CFB mainly introduces opportunities for \emph{previewing the topical content of videos}, more emphasis is given to capturing activities related to users exploring topics in the videos.} The interaction metric categories that we analysed, with or without CFB, were as follows:

\paragraph{Time} \textcolor{black}{We computed a set of interaction metrics that are associated to the time participants spent on different actions. The metrics that we analysed were: i) Time spent completing the tasks in the study, ii) time spent in the results screen, iii) time spent watching the videos during the task, iv) time spent exploring the content of videos during the task, v) time spent exploring the contents of videos in the results screen, vi) time spent exploring the content of videos in the video player screen, vii) time spent watching the video per opened video, and viii) time spent exploring the content of videos per explored video. When considering exploration of video content, we measured the time spent hovering over the time series bar (as explained in section \ref{sec:user_actions}).}

\paragraph{Activity} \textcolor{black}{To analyse different patterns of activity, we measured seven interaction metrics that fall under frequency and proportion based metrics. These metrics are: i) Unique number of videos opened/played, ii) number of videos played, iii) number of play sessions per unique video played (i.e. fraction of (i)/(ii)), iv) number of segments removed within 1 minute of selection, v) number of segments removed, vi) time spent exploring in result screen as a fraction of entire task duration and vii) time spent exploring in video player screen as a fraction of entire task duration.}

\paragraph{Navigation} \textcolor{black}{We computed several metrics that gave us cues about where participants navigate within the system. These were: i) Number of seek actions through the entire task (a seek action is when a user jumps to a different time locations in the video to play), ii) number of seek actions within a video per opened video, iii) deepest rank of videos played (i.e. the maximum rank in the list of the videos opened), iv) deepest rank of the video explored (i.e. similar to the previous metric but for videos explored), and v) mean position navigated within the video (i.e. the average position in the video, normalised in [0,1], that the participants watched. The deepest rank gave us insights regarding whether  users were encouraged to consider results that are in lower ranks in the results screen. Larger numbers indicate that users' attention was caught by lower ranked items.}

\paragraph{Selection} \textcolor{black}{We computed some further metrics related to the segments participants selected as relevant to their task. These were the following: i) Time spent before first selection, ii) number of videos opened/played before selecting the first segment, iii) number of videos explored before selecting the first segment 
and iv) average duration of selected segments.}

\begin{table*}[]
\caption{Various metrics calculated using the interaction logs recorded during the user study presented with the mean and the statistical analysis done on 5 random user pairing configurations (mean pairwise difference between CFB off and CFB on). The differences in pairs that are statistically significant are marked with $p<0.01$ as $^{(***)}$, $p<0.05$ as $^{(**)}$ and $p<0.10$ as $^{(*)}$.}
\label{tab:results}
\scriptsize
\centering
\begin{tabular}{llcc}
\toprule
 &  & \multicolumn{1}{c}{Mean Pairwise Difference (\texttt{CFB On - CFB Off})} & \multicolumn{1}{c}{Mean Wilcoxon p value} \\
 \midrule
 & Time spent completing the task & -48.607 & 0.8034 \\
 & Time spent in results screen & 168.267 & 0.0305$^{(**)}$ \\
 & Time spent watching videos during the task & -136.867 & 0.3718 \\

Time & Time spent exploring during the task & 189.447 & 0.0148$^{(**)}$\\
(secs.) & Time spent exploring in results screen & 139.160 & 0.0004$^{(***)}$\\
 & Time spent exploring in video player screen & 50.287 & 0.3005\\
  & Time spent watching videos per opened video & -22.296 & 0.5338\\
 & Time spent exploring in results screen per explored video &  16.524 & 0.0082$^{(***)}$\\
 \midrule
 & Number of unique videos played & 0.020 & 0.8748\\
& Number of videos played &-0.507 & 0.5566\\
& Number of play sessions per unique video played & -0.018 & 0.6416\\
Activity & Number of segments removed within 1 minute of selection & -0.113 & 0.1528\\
 & Number of segments removed & -0.113 & 0.3885\\
& Fraction time spent exploring in result screen per task & 0.132 & 0.0005$^{(***)}$\\
& Fraction time spent exploring in video player screen per task & 0.082 & 0.1795\\
 \midrule
 & Number of seek actions & -8.167 & 0.2899\\
 & Number of seek actions per played video & -1.465 & 0.4080\\
Navigation & Deepest rank of video played & -0.153 & 0.6913\\
 & Deepest rank of video explored & 1.873 & 0.0336$^{(**)}$\\
 & Mean position played within the video (fraction) & 0.036 & 0.5555\\
  \midrule
& Time spent before first selection & -13.867 & 0.7122\\ 
Selection & Number of videos played before first selection & -0.533 & 0.0642$^{(*)}$\\
& Number of videos explored before first selection & -0.193 & 0.2903\\
 & Average duration of selected segments & -12.841 & 0.8042\\
 \bottomrule
\end{tabular}
\end{table*}

\subsection{Statistical Analysis} \label{sec:stat_an}

One of the goals of the user study is to objectively compare the interaction metrics for the two conditions (CFB on vs. off) to see if the distributions from the two groups are statistically different. 
Many metrics from interacting with web-based documents (e.g. watch time for videos \cite{context_agnostic_engagement}, dwell time with clicks \cite{beyond_clicks}) tend to follow distributions that are non-normal, motivating instead the use of non-parametric hypothesis tests \cite{Guo_vid_prod}. Due to the non-normality in our measurements, we used the \emph{Wilcoxon signed-rank test} \cite{conover1998practical} to compare the difference in values between the two conditions 
. 

\textcolor{black}{As per section \ref{sec:experiment}, two very different knowledge areas (climate change and machine learning) were used for the two conditions to avoid the learning effect. Our initial analysis showed that there are significant behavioral differences between the two topics, possibly due to the different videos and topics contained or the background of the participants (see section \ref{sec:task_diff} and \figurename{ \ref{fig:boxplots} (c)}). Therefore, exclusively comparing the same participant's two conditions as a paired observation can be misleading. To address this issue, we paired the control task (CFB off) of each participant with a randomly chosen different participant who carried out the treatment task (CFB on) for the same topic. Once pairing is done, the new paired observations were used to run Wilcoxon signed-rank test. To assure statistical reliability \textcolor{black}{of the result obtained by synthetically pairing observations, random pairing is iterated 5 times.} The results of hypothesis testing done on the 5 participant pairing configurations is reported in Table \ref{tab:results}.} \textcolor{black}{Detailed results obtained from the 5 pairing configurations is attached with the supplementary materials.}


\section{Effect of CFB on user interaction behaviours and perceptions}

\textcolor{black}{We focus our analysis mostly on the results obtained via statistical analysis of the user interaction logs, 
the survey questionnaire
and the thematic analysis. Table \ref{tab:results} presents the outcome of the statistical analysis conducted on the interaction logs. We comment both on the mean pairwise difference and the significance of the statistical tests. \figurename{ \ref{fig:boxplots}} contains several box plots that indicate noteworthy differences of behaviours between tasks. \figurename{ \ref{fig:barplogquestionnaire}} summarises key results obtained by analysing the questionnaire.}

\subsection{Differences Across Tasks} \label{sec:task_diff}

    \begin{figure*}[ht!]
    \centering
    \includegraphics[width=\textwidth]{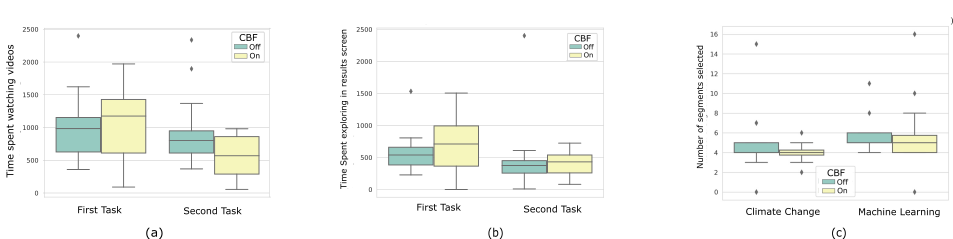}
    \caption{
    Box plots of a subset of interaction metrics indicating differences of interaction behaviours between task order and topic. These plots (specifically subfigure a) and b)) suggest a learning effect taking place where users spend much longer time on the first task. Subfigure c) shows that the number of segments selected for the two different tasks to be different although the users were encouraged to spend similar time periods and to select the same number of segments for both tasks indicating different complexities of the two topics.
    }
    \label{fig:boxplots}
    \end{figure*}
    

\textcolor{black}{\figurename{ \ref{fig:boxplots}} shows three box plots that summarise a relevant finding. In summary, we found marked differences in the interaction log results when taking into account task topic and task order. Subfigure a) and b) in \figurename{ \ref{fig:boxplots}} indicate that the time taken to perform the first task is systematically larger than the time taken to perform the second task regardless of the condition (CFB on/off). This suggests that there may be a learning effect taking place during the study. This learning effect in the first task seems to be more pronounced when participants are exposed to the CFB. In the second task, however, both conditions take similar time, with CFB in some cases taking less time. This may indicate that the visual CFB tool is intuitive, since the overall task duration of the second task converges to the same amount than that of the group that had CFB disabled. We believe this is evidence that users can understand and learn the CFB functionality by interacting with it for less than one hour. Subfigure c) shows that the number of segments selected  by the users is different based on the topic of the task although the users were advised to select the same number of segments at the end of each task. This may be an indication of difference of complexity of the task that is associated with the topic, \textcolor{black}{since participants were more familiar with the topic of machine learning. Additionally, although the tasks are similar, the climate change one is more specific, asking participants to look for i) applications of data science to climate change and ii) controversial issues in climate change, whereas the machine learning task simply asks to highlight relevant material for understanding the topic and showing its applications.}}

These observations motivated us to perform the statistical analysis (outlined in section \ref{sec:stat_an}) by pairing user tasks that are similar in task order and task topic rather than using the individuals two conditions as a paired observation.

\subsection{Time Spent}




\textcolor{black}{
The main conclusion from the time metrics in \tablename{ \ref{tab:results}} is that
participants spend much more time choosing and exploring what content to open and watch when using CFB. This suggests that users may be utilising the CFB to locate topical content across the videos.
This is evidenced by the significance of the tests of metrics such as time spent in the results screen, time spent exploring during the task and time spent exploring in the results screen.
This behaviour may suggest that they are using the keywords to make decisions about which videos to watch. 
This may indicate that there is promise in CFB enabling users to navigate within a video collection.}

\textcolor{black}{
Although the results are not statistically significant, the time metrics in \tablename{ \ref{tab:results}} also show that participants spend less time completing the task with CFB. They also spend less time watching material. Finally, the results also show that participants spend more time exploring the content within a specific video.}

\subsection{Users Activity}

\textcolor{black}{
The activity group shows again that participants spend most of the time during completion of the task exploring videos in the result screen. Although not statistically significant, they also spend more time exploring in the video player screen. However, they watch less videos on average, and have less play sessions per unique video played. This may indicate they can make better or quicker relevance judgments.
The {number of segments removed within 1 minute of selection (and across the study)} is also less when using CFB.
 Such behaviour is analogous to "quickback" behaviour in information retrieval when a user reverts their decision, usually caused by a mistake action. This suggests that having CFB leads to less mistakes in terms of selecting relevant sections.}

\subsection{Users Navigation}

\textcolor{black}{The navigation metrics in \tablename{ \ref{tab:results}} give a strong indication that the CFB enables users to explore deeper in the result set.  This evidence demonstrates the utility of this user interface component to support users even in systems where the ranking algorithms may not be perfect. 
Although not statistically significant, the deepest rank of video played is less with CFB. This may suggest that, with CFB, participants may have found evidence of relevance earlier in the result set.}
\textcolor{black}{Additionally, \tablename{ \ref{tab:results}} also shows that the number of seek actions performed by users is less with CFB. This suggests that the tool also empowers users to navigate within a video. However, the effect is not as strong as found in the results screen.
These results lead us to hypothesize that using the CFB, participants explore more videos, but also spend more time checking the topical content in those (as suggested by the conclusions in the previous subsection). However, once they decide to open a material, they can decide better where to go, and need to take less seek actions within the video. This is also evidenced by the mean position played within the video, which shows that with CFB participants play deeper in the video, even though on average they watch less material. }

\subsection{Selecting Relevant Content}


\textcolor{black}{The selection metrics in \tablename{ \ref{tab:results}} indicate that participants using CFB play less number videos before selecting the first relevant video segment for the task. On average, participants also explore less videos before making a relevance judgment and they also spend less time before selecting a relevant segment of the video. 
Additionally, the average duration of selected segments shows that participants with CFB select less content. We believe that this is because they can select much more fine-grained segments. }

\subsection{Users Perception of the CFB}

\begin{figure}[ht!]
    \centering
    \includegraphics[width=0.5\textwidth]{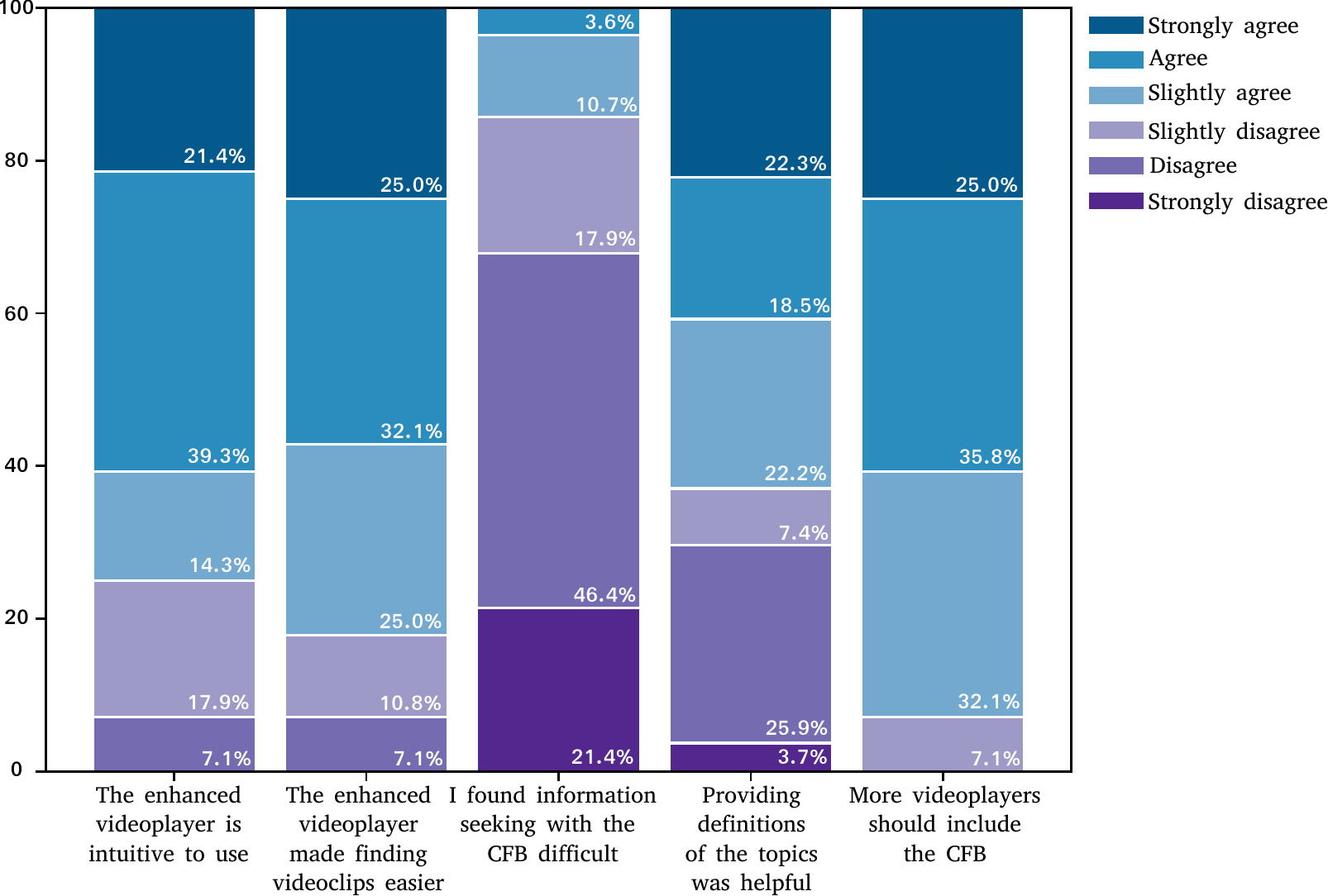}
    \caption{Responses to a selection of questions from the user questionnaire. A 6-grade Likert scale is used. Different shades of blue are used for agreement responses, while different shades of purple are using for disagreement responses.}
    \label{fig:barplogquestionnaire}
    \end{figure}

\paragraph{Questionnaire Survey}
\textcolor{black}{From the questionnaire rating scales, we can see (see \figurename{ \ref{fig:barplogquestionnaire}}) that most participants had a positive reaction to the CFB (82\% agreed it made finding videoclips easier). The majority agreed that they enjoyed using the CFB and that
it was intuitive (75\% agreed). Despite some issues reported by participants with the adequateness of some keywords, participants seemed to be satisfied with the CFB. Almost 93\% agreed that more videoplayers should include the CFB.} 

\paragraph{Thematic Analysis}
\textcolor{black}{
When analysing the responses to the open ended questions in the questionnaire using thematic analysis, we found two emerging main themes. These were i) supported exploration of video content and ii) positive user experience. We demonstrate these now by including some of the statements made by participants.
Similarly to the results in the user survey, participants were positive about the tool, one stated \emph{"I really like that feature, I would use it often"}. 
 Additionally, the CFB seemed helpful and easy to use for users. \emph{"Once you explained it in a couple of minutes, it would be quite clear to most people how to use it."}  \emph{“Overall, the flowbar made searching feel much more focused.”} \emph{“It is really helpful that many terminologies have explanations of them when you hover over them”.} The reaction of some of the participants when trying to select video clips without the CFB (especially in long videos) is noteworthy. \emph{"Once I got to the machine learning, when you said find 4 clips...urg"}. Specifically, participants seemed frustrated with the effort required to select the clips without the CFB: \emph{"Definitely, I think you saw I got quite frustrated in the first one with the standard video player without the keywords"}.}
\textcolor{black}{Most participants highlighted several advantages of the CFB regarding exploring video content in the open-ended questions, which can be summarized as: i) enhancing video navigation and locating information, and ii) getting an overview of the videos. 
Participants could more easily scroll and skim through the video, which was especially important for lengthy ones. 
\emph{“It helps to skip-read the content of the video, especially for the long videos”}.
They could locate information more quickly and it enabled them to select directly which parts of videos they wanted to look at.
\emph{“If you have a fixed amount of time you can get through a greater amount of material”}.
Participants looked first at thumbnails and video titles, then the CFB keywords provided extra cues, which let them select directly the video segments that they wanted to look at. \emph{“I sweep straight forward starting from the section rather than the whole thing just clicking randomly through the video”}.  If participants could not find directly a segment that seemed relevant, the CFB then helped them to narrow down their search and eliminate information.
\emph{“It made identifying the detail of the subject being spoken about much easier; it gave some sense of where the speaker was in the flow of conversation.”}
\emph{“CFB helps to find starting points within the videos, not necessarily always helping me to find relevant content actually, but at least to quickly move through.”}
The CFB enabled participants to  get an overview of the video content, and enhance their understanding of the video. \emph{"It increased my knowledge about the video"}. It gave participants a digest of each video and a quick summary of different video sections. It gives them some context for the ideas presented, allowing them to get a much better sense of the flow of the videos. As stated: \emph{"These keywords can help in summarizing what the idea is about and differentiating one".} 
}

\textcolor{black}{Despite participants positive reactions to the CFB, they also noted a few issues. A few misappropriate or misleading keywords were generated. Most participants seem to have ignored them, although a few were puzzled when they were noticeably out of context and/or incongruous. Some participants highlighted that there is a lot of repetition across the keywords, thus making the keywords less salient.  \emph{"For some videos, the keywords were the same for several segments making it difficult to distinguish between them"}. In particular, for a few videos, a keyword was repeated systematically across the video. As a participant pointed such keywords should be left out, and included as meta-tag for the whole video.} 
\textcolor{black}{However, the reaction from the participants also provided assurance that Wikification, despite its occasional errors, improved navigation between and within videos. This could be supporting evidence of the utility of automatic and scalable entity linking for such content navigation tools.}




\section{Discussion and Conclusions}


We presented the design rationale for the Content Flow Bar (CFB) and the analyses from our user study describing the efficacy of this novel interaction technique for supporting discoverability and navigation. Overall, the CFB was found to enhance early exploratory behaviour and information seeking across sets of videos. It does this by providing visual cues about the different topics covered throughout a video that otherwise may not have been perceptible. \textcolor{black}{Two relevant findings from both the user interaction logs with the tool and the questionnaire was that participants were able to explore more materials with the CFB and found the tool helpful for information retrieval tasks (e.g. watched less videos before making a relevance judgment).}

\textcolor{black}{
Due to the low number of participants in our user study (26), not all the results in the interaction log analysis are significant. However, we also analysed the means with and without CFB to extract conclusions regarding this new tool. The findings were as follows: 
Participants seemed to spend less time on average to complete the tasks using CFB. The distribution of time was also different. With CFB, they spent much more time deciding on what content to watch and also from which starting point. This is, they explored more across sets of videos and within a specific video. On the other hand, participants spent on average less time watching material with CFB, and they opened less videos. They also made less mistakes in relevance judgments when using CFB (as evidenced by the number of segments marked as relevant and then removed by participants). Participants also explored deeper in the result list and seeked less within a video with CFB. Finally, it seemed that participants took less time with CFB to make relevance judgments and they selected more fine-grained content segments.
The questionnaire highlighted that most participants had a positive experience with the CFB. They found the tool intuitive and helpful in providing summaries of the content and helping them in information seeking tasks.}

The CFB tool is intended to be agnostic of content type and scalable to large collections of documents, with the potential of being used automatically across different modalities of informational content (e.g. videos, pdfs, audio, etc.). That is, while many search tools rely on manual annotations or are specific to particular content type, there is a scarcity of generic, scalable and automatic solutions for users to preview and search across diverse modalities of content. The backbone of our interaction tool (Wikification), allows for such features. We hypothesize that the benefits found may go beyond educational videos and can be used for a wide range of long document formats, potentially making it a powerful tool in the hands of learners and teachers. 

In future work, we plan on conducting a more extensive analysis of the interaction logs and user experiences. 
\textcolor{black}{For example, we will consider subject expertise as a covariate in the data analysis to confirm whether it has an impact on users information seeking and exploration behavior. We will also analyse and compare the quality of selected segments between the systems through a pairwise comparison experiment. This will help us to further investigate the usefulness of the tool for relevance judgments.}
\textcolor{black}{We propose that running a larger user study comprising of a more diverse group of users, a larger set of information retrieval and navigation tasks, additional content modalities (apart from video) and other more advanced control conditions (for example, a video player that would allow users to search in the video transcript), is needed to evaluate even further the potential of such content navigation tools for search and exploration.} 


\textcolor{black}{In subsequent versions of the CFB, we plan on addressing the limitation of having fixed size fragments in entity linking and thus the CFB. This is, we aim to dynamically infer where topics change and have variable size fragments.
Additionally, we hypothesize that ultimately keywords should be personalised, based on the background knowledge and goals of the user, showing more general topics to novice users and more specific concepts to advanced learners, and highlighting parts of the content that match the timely goals of the user.
We also aim to explore other designs of the tool that allow to provide hierarchies of the topics covered in the material, as well as visualisations that help the user to further understand the 'content flow' of the material. 
A major limitation of the presented CFB tool and the two visualisations shown is that they are reliant on transcription, translation and wikification algorithms, which as we discovered are still prone to errors. However, as these improve, navigation and visualisation tools like CFB will also improve, and in so doing, are likely to prove very useful for informational search tasks. In particular, as our study has shown, they can help users find relevant online material rapidly rather than having to spend ages sifting through a multitude of search results to find what they are looking for. In sum, they enable learners and educators "to watch less and uncover more" when searching and exploring educational material.  }

\begin{acks}
This research was partially conducted as part of the X5GON project funded from the EU's Horizon 2020 research programme grant No 761758.
This work is also supported by the European Commission funded project "Humane AI: Toward AI Systems That Augment and Empower Humans by Understanding Us, our Society and the World Around Us" (grant 820437) and the EPSRC Fellowship titled "Task Based Information Retrieval" (grant EP/P024289/1). We also thank Shenal Pussegoda for his  help and assistance with the implementation of the tools. 
\end{acks}








\end{document}